\documentclass[notitlepage,nofootinbib,aps,prd,showpacs,onecolumn,
superscriptaddress,article]{revtex4-1}

\usepackage{graphicx}
\usepackage{dcolumn}
\usepackage{amsmath}
\usepackage[makeroom]{cancel}
\usepackage{amssymb}
\usepackage{bm}
\usepackage{color}
\usepackage[normalem]{ulem}
\usepackage[dvipsnames]{xcolor}
\usepackage{hyperref}
\usepackage[many]{tcolorbox}

\hypersetup{
colorlinks=true, 
linkcolor=blue,  
citecolor=cyan,  
}
\usepackage{tikz}
\makeatletter
\newcommand*{\encircled}[1]{\relax\ifmmode\mathpalette\@encircled@math{#1}\else\@encircled{#1}\fi}
\newcommand*{\@encircled@math}[2]{\@encircled{$\m@th#1#2$}}
\newcommand*{\@encircled}[1]{%
  \tikz[baseline,anchor=base]{\node[draw,circle,outer sep=0pt,inner sep=.2ex] {#1};}}
\makeatother

\usetikzlibrary{arrows}
\usetikzlibrary{shapes}
\newcommand{\mymk}[1]{%
  \tikz[baseline=(char.base)]\node[anchor=south west,color=red, draw,rectangle, rounded corners, inner sep=2pt, minimum size=7mm,
    text height=5mm](char){\ensuremath{#1}} ;}

\newtcolorbox{cross}{blank,breakable,parbox=false,
  overlay={\draw[blue,line width=2pt] (interior.south west)--(interior.north east);
    \draw[blue,line width=2pt] (interior.north west)--(interior.south east);}}

\begin{document}
%
\title{Raychaudhuri equations and gravitational collapse in Einstein--Cartan theory}

\author{Sudipta Hensh}
\email{f170656@fpf.slu.cz, sudiptahensh2009@gmail.com}
\affiliation{Research Centre for Theoretical Physics and Astrophysics, Institute of Physics, Silesian University in Opava, Bezru\v{c}ovo n\'{a}m\v{e}st\'{i} 13, CZ-74601 Opava, Czech Republic}

\author{Stefano Liberati}
\email{liberati@sissa.it}
\affiliation{SISSA - International School for Advanced Studies, Via Bonomea 265, 34136 Trieste, Italy }
\affiliation{IFPU - Institute for Fundamental Physics of the Universe, Via Beirut 2, 34014 Trieste, Italy }
\affiliation{INFN Sezione di Trieste, Via Valerio 2, 34127 Trieste, Italy }

%

%
\date{\today}
\begin{abstract}
The Raychaudhuri equations for the expansion, shear and vorticity are generalized in a spacetime with torsion for timelike as well as null congruences. These equations are purely geometrical like the original Raychaudhuri equations and could be reduced to them when there is no torsion. Using the Einstein--Cartan--Sciama--Kibble field equations the effective stress-energy tensor is derived. We also consider an Oppenheimer--Snyder model for the gravitational collapse of dust.  It is shown that the null energy condition~(NEC) is violated before the density of the collapsing dust reaches the Planck density, hinting that the spacetime singularity may be avoided if there is a non-zero torsion, i.e.~if the collapsing dust particles possess intrinsic spin.

\end{abstract}


\maketitle

\section{Introduction}
%


The evolution of a congruence in a spacetime is determined by the so called Raychaudhuri equation~\cite{Raychaudhuri}. An important fact about those equations is that they are purely geometrical and do not assume any theory of gravity. This feature gives freedom to use any theory to fix the geometry and then study the evolution of congruence. Its importance was realized greatly upon its use in establishing the so called Hawking--Penrose singularity theorem~\cite{Penrose1965,Hawking1965,Hawking1966}. 

Einstein's theory of gravity considers spacetime on Riemannian manifolds which assumes vanishing torsion and zero non-metricity. The most general spacetime can be found relaxing these assumptions. The generalization of Einstein's gravity in a spacetime with torsion is known as Einstein--Cartan theory~(ECT). The origin of this torsion in ECT is due to intrinsic spin of elementary particles. The field equations for ECT were found by Sciama~\cite{Sciama1962} and Kibble~\cite{Kibble1961} independently  in the 60's. Remarkably outside matter distributions the geometry is completely determined by Einstein's general relativity~(GR) due to the non-propagating nature of torsion in ECT.

To understand the evolution of congruences in ECT it is necessary to generalize the Raychaudhuri equation in a spacetime with torsion. These can be found in some articles in the literature~\cite{Capozziello2001,Kar2007,Wanas2009,Cai2016,Luz2017,Pasmatsiou2017} but not for full set of $N$-dimensional Raychaudhuri equations for expansion, shear and vorticity considering both timelike and null congruences as we present it here. While we cannot do justice in mentioning all efforts available in the literature, we shall try at least to mention the most relevant ones for this work. In particular, the generalization of the Raychaudhuri equations in the presence of torsion was studied in~\cite{Luz2017,Kar2007,Wanas2009,Cai2016,Capozziello2001}. In the most general case, i.e.~considering also non-zero non-metricity, the Raychaudhuri equation was found in~\cite{Iosifidis2018}. Null geodesic congruences in the presence of torsion were studied in~\cite{Speziale2018}. In~\cite{Dey2017}, the properties of Killing horizon in the spacetime with torsion were investigated. 

In stellar objects evolution, the degenerate pressure due to Pauli exclusion principle may replace thermonuclear fusion in counterbalancing the inward self gravity, leading in this case to white dwarfs or neutron stars. However, for sufficiently massive objects this is not possible so that gravitational  collapse cannot be avoided and eventually possibly leading to black hole formation. Oppenheimer and Snyder (OS)~\cite{Oppenheimer1939} studied the gravitational collapse of pressureless dust, i.e.~matter formed by non-interacting particles characterized by non-zero mass-density but negligible pressure.\footnote{A generalization of Oppenheimer and Snyder solution was found in close analytical form by P.C. Vaidya in~\cite{Vaidya1966} where he considered the collapsing dust with radiation flowing outward.} While this might seem to be a very idealistic model, its analytical form is sometimes very helpful in gaining an intuitive  understanding of the most relevant physical effects at work and for this reason we shall use it here for a first hand exploration about the possible effects on torsion.

One of requirements for Penrose's singularity theorem~\cite{Penrose1965} to hold is that the null energy condition~(NEC) --- $T_{\mu\nu}k^\mu k^\nu\geq 0$ for any null vector $k^\mu$ --- should always be satisfied during the gravitational collapse. In this sense, the violation of NEC at some point of the collapse indicates that the formation of a spacetime singularity might be avoided. Indeed, we shall see that the presence of torsion may cause such a violation after the formation of a trapping horizon but well before a Planckian (quantum gravity) regime is reached. Although this is projected to happen at very high densities, at which we do not have yet a full understanding of matter behaviour, it as well true that the latter is not expected at these late stages to be dominant over gravitational effects. Also our results lends support to previous investigations on the possible role of torsion in avoiding singularities.  For example, an intuitive argument on this can be found in~\cite{Majhi2018} within the asymptotically safe gravity framework. Similarly, it was shown that the cosmological singularity might be avoided in ECT~\cite{Trautman1973}. Also OS collapse in ECT was numerically studied finding that the singularity formation is avoided by a bounce~\cite{Hashemi2015}. Our analytical study can be considered as complementary to this last one.

The paper is structured as follows: after setting our conventions and notation in Section~\ref{def}, we generalize in Section~\ref{sec:evol} to spacetimes with torsion the usual description of the evolution of the separation vector for curves in a congruence.  In Section~\ref{timelike} we then derive the Raychaudhuri equations for expansion, shear and vorticity for a timelike congruence.  In Section~\ref{null} we derive the same equations for a null congruence. After reviewing in Section~\ref{sec:ECSK} we review the Einstein-Cartan--Sciama--Kibble field equations, we then discuss the OS collapse in the presence of torsion in Section~\ref{sec:ECcoll}. In Section~\ref{sec:disc} we summarize our results and discuss future perspectives.

Finally, note that while in the first part of the paper, i.e.~in deriving the Raychaudhuri equations for spacetime with torsion, we take $G = c=1$, in the second part of the manuscript, concerning the OS collapse model in ECT, we restore the actual values of these constants so to provide numerical estimates for the relevant physical quantities at play.
Throughout the manuscript we use the signature $(-,+,+,+)$, Greek indices run from $0$ to $3$ while Latin ones run from $1$ to $3$.

\section{Definitions and notations \label{def}}
The covariant derivative for a generic four-vector~$X^{\beta}$ is defined as
\begin{equation} \label{eq:1}
\nabla_{\alpha} X^{\beta} = \partial_{\alpha} X^{\beta} + C^{\beta}_{\,\,\,\alpha \sigma} X^{\sigma} \ .
\end{equation}
In our considerations the connection, $C^{\gamma}_{\,\,\,\alpha \beta}$ will have the only constrain of being metric compatible i.e. 
\begin{equation} \label{eq:2}
\nabla_{\alpha} g_{_{\beta \gamma}} = 0  \quad (\textrm{zero non-metricity}) \ .
\end{equation}
The torsion tensor is defined as the anti-symmetric part of the generic connection which is given by
\begin{equation} \label{eq:3}
{S_{\alpha \beta}}^{\gamma} \equiv C^{\gamma}_{\,\,\,[\alpha \beta]} = \frac{1}{2} \left( C^{\gamma}_{\,\,\,\alpha \beta} - C^{\gamma}_{\,\,\,\beta \alpha} \right) \ .
\end{equation}
${S_{\alpha \beta}}^{\gamma}$ is antisymmetric in its first two indices i.e.
\begin{equation} \label{eq:4}
{S_{\alpha \beta}}^{\gamma} = - {S_{\beta \alpha}}^{\gamma} \ .
\end{equation}
The general metric compatible connection can be written as the addition of Levi--Civita~($\Gamma^{\gamma}_{\,\,\,\alpha \beta}$) connection and contorsion tensor~(${K_{\alpha \beta}}^{\gamma}$),
\begin{equation} \label{eq:5}
C^{\gamma}_{\,\,\,\alpha \beta} = \Gamma^{\gamma}_{\,\,\,\alpha \beta} + {K_{\alpha \beta}}^{\gamma} \ ,
\end{equation}
where ,
\begin{equation} \label{eq:6}
{K_{\alpha \beta}}^\gamma \equiv {S_{\alpha \beta}}^\gamma + {S^\gamma}_{\alpha \beta} - S_{\beta \,\,\,\alpha}^{\,\,\,\gamma} \ .
\end{equation}
Let us now consider the Lie derivative of a vector $v$ along another vector $u$ takes the form,
\begin{eqnarray} \label{eq:7}
\mathcal{L}_u v \equiv [u,v]^\gamma = u^\alpha \partial_\alpha v^\gamma - v^\alpha \partial_\alpha u^\gamma \ ,
\end{eqnarray}
we can see that by using Eq.~\eqref{eq:1}, the expression in Eq.~\eqref{eq:7} can be written as,
\begin{eqnarray} \label{eq:8}
\mathcal{L}_u v = u^\alpha \nabla_\alpha v^\gamma - v^\alpha \nabla_\alpha u^\gamma -2 {S_{\alpha \beta}}^{\gamma} u^\alpha v^\beta \ ,
\end{eqnarray}
where we use the definition of torsion tensor given in \eqref{eq:3}.

The definition of Riemann tensor is given by,
\begin{equation} \label{eq:9}
{R_{\alpha \beta \gamma}}^ \rho = \partial_\beta C^{\rho}_{\,\,\,\alpha \gamma} - \partial_\alpha C^{\rho}_{\,\,\,\beta \gamma} + C^{\rho}_{\,\,\,\beta \sigma} C^{\sigma}_{\,\,\,\alpha \gamma} - C^{\rho}_{\,\,\,\alpha \sigma} C^{\sigma}_{\,\,\,\beta \gamma} \ .
\end{equation}
The commutator of covariant derivative reads,
\begin{eqnarray} \label{eq:10}
[\nabla_\alpha, \nabla_\beta] w_\gamma &=& {R_{\alpha \beta \gamma}}^ \rho w_\rho - 2 {S_{\alpha \beta}}^{\rho} \nabla_\rho w_\gamma \ ,
\end{eqnarray} 
here we use the commutativity of the partial derivative.
Similarly, the generalized Ricci tensor is
\begin{equation} \label{eq:11}
R_{\alpha \beta} \equiv {R_{\alpha \gamma \beta}}^ \gamma = \partial_\gamma C^{\gamma}_{\,\,\,\alpha \beta} - \partial_\alpha C^{\gamma}_{\,\,\,\gamma \beta} + C^{\rho}_{\,\,\,\alpha \beta} C^{\gamma}_{\,\,\,\gamma \rho} - C^{\rho}_{\,\,\,\gamma \beta} C^{\gamma}_{\,\,\,\alpha \rho} \ ,
\end{equation}
and the Ricci scalar is defined as usual,
\begin{equation} \label{eq:12}
R \equiv g^{\alpha \beta} R_{\alpha \beta} \ .
\end{equation}

\section{Evolution of separation vector}\label{sec:evol}
In this section we review the formalism introduced in ref.~\cite{Luz2017} for describing the calculation of the separation vector among curves of a congruence.
Let us consider a congruence $\gamma_s (\lambda)$ where $s$ changes from one curve to another and $\lambda$ changes along the curve.
Let us choose two points $p$ and $q$ lying on two adjacent curves having coordinates 
${x^\alpha(\lambda)}$ and ${x'^\alpha(\lambda)} = {x^\alpha(\lambda) + \xi^\alpha}$ respectively.
Here $\xi^\alpha$ is the separation vector is given by,
\begin{eqnarray} \label{eq:13}
\xi^\alpha = \frac{\partial x^\alpha}{\partial s} \ .
\end{eqnarray}
The tangent vector along the curve is defined as
\begin{eqnarray} \label{eq:14}
u^\alpha = \frac{\partial x^\alpha}{\partial \lambda} \ .
\end{eqnarray}
In these coordinates the Lie-derivative of the tangent vector along the separation vector (and the vice versa) $\xi^\alpha$ is trivially null
\begin{eqnarray} \label{eq:15}
\mathcal{L}_\xi u &=& 0 = \mathcal{L}_u \xi \ .
\end{eqnarray}
%
Using~\eqref{eq:8} and~\eqref{eq:15} we then get,
\begin{eqnarray} \label{eq:16}
u^\beta \nabla_\beta \xi^\alpha &=& {B_\beta}^\alpha \xi^\beta ,
\end{eqnarray}
where $B_{\alpha \beta}$ is given by,
\begin{equation} \label{eq:17}
B_{\alpha \beta} \equiv \nabla_\alpha u_{\beta} + 2 S_{\gamma \alpha \beta} u^\gamma \ .
\end{equation}

In general, $B_{\alpha \beta}$ can be decomposed in orthogonal and parallel components to the congruence
\begin{equation} \label{eq:18}
B_{\alpha \beta} = B_{\perp \alpha \beta} + B_{\parallel \alpha \beta} \ .
\end{equation}
Defining as usual the projection operator as
$h_{\alpha \beta} = g_{\alpha \beta} - \sigma u_\alpha u_\beta$
where $\sigma$ is equal to -1 or 1 depending on whether the tangent vector is timelike or spacelike,
we can write these components as
\begin{eqnarray}
B_{\perp \alpha \beta} &\equiv& h^{\,\,\,\gamma}_\alpha h^{\,\,\,\sigma}_\beta B_{\gamma \sigma} \label{eq:19} \\
B_{\parallel \alpha \beta} &\equiv& B_{\alpha \beta} - B_{\perp \alpha \beta} \label{eq:20} \ .
\end{eqnarray}
Expansion, shear and vorticity are then defined as
\begin{eqnarray}
\theta &=& {B_{\perp \gamma}}^\gamma \ , \label{eq:21} \\
\sigma_{\alpha \beta} &=& B_{\perp (\alpha \beta)} - \frac{h_{\alpha \beta}}{h^{\,\,\,\gamma}_\gamma} \theta \ , \label{eq:22} \\
\omega_{\alpha \beta} &=& B_{\perp [\alpha \beta]} \label{eq:23}\ ,
\end{eqnarray}
so that $B_{\perp \alpha \beta}$ can be decomposed as
\begin{equation} \label{eq:24}
B_{\perp \alpha \beta} = \frac{h_{\alpha \beta}}{h^{\,\,\,\gamma}_\gamma} \theta + \sigma_{\alpha \beta} + \omega_{\alpha \beta} \ .
\end{equation}

\section{Raychaudhuri equation with torsion for timelike congruences \label{timelike} }
Let us now consider timelike congruences. The projection operator in this case is 
\begin{equation} \label{eq:25}
h_{\alpha \beta} \equiv g_{\alpha \beta} + v_\alpha v_\beta \ ,
\end{equation}
and satisfies
\begin{eqnarray} \label{eq:26}
h_{\alpha \beta} v^\alpha = 0 \ , \quad
h^{\,\,\,\gamma}_\alpha h_{\gamma \beta} = h_{\alpha \beta} \ , \quad
h^{\,\,\,\gamma}_\gamma = N-1 \ ,
\end{eqnarray}
where $N$ is the dimension of the spacetime. Using the definitions given in Eq.~~\eqref{eq:19} and Eq.~\eqref{eq:25}, we get

\begin{eqnarray} \label{eq:27}
B_{\perp \alpha \beta} &=& \nabla_\alpha v_\beta + 2 S_{\rho \alpha \beta} v^\rho + 2 S_{\rho \alpha \sigma} v^\rho v^\sigma v_\beta + v_\alpha a_\beta  \ ,
\end{eqnarray}
where
$a_\beta = v^\gamma \nabla_\gamma v_\beta$ . Here we used the metric compatibility and the contraction of antisymmetric (torsion) and symmetric tensor ($v^\alpha v^\beta$) is zero.
Rewriting Eq.~\eqref{eq:27} in terms of Levi--Civita derivatives one gets
\begin{eqnarray} \label{eq:34}
B_{\perp \alpha \beta} &=& \widetilde{\nabla}_\alpha v_\beta 
- {K_{\alpha \beta}}^\sigma v_\sigma
+ 2 S_{\rho \alpha \beta} v^\rho 
+ 2 S_{\rho \alpha \sigma} v^\rho v^\sigma v_\beta 
- {K_{\rho \beta}}^\sigma  v^\rho v_\sigma v_\alpha \ .
\end{eqnarray}
From the definitions Eq.~\eqref{eq:20} and Eq.~\eqref{eq:17}, it then follows
\begin{eqnarray} \label{eq:28}
B_{\parallel \alpha \beta} &=& - 2 S_{\rho \alpha \sigma} v^\rho v^\sigma v_\beta - v_\alpha a_\beta \ .
\end{eqnarray}
The perpendicular component of $B_{\alpha \beta}$ without torsion can be written as
\begin{eqnarray} \label{eq:29}
\widetilde{B}_{\perp \alpha \beta} &=& \widetilde{\nabla}_\alpha v_\beta \ ,
\end{eqnarray}
where $\widetilde{\nabla}$ is the covariant derivative w.r.t. Levi--Civita connection.

\subsection{Raychaudhuri equation for the congruence expansion}

Using~\eqref{eq:29}, the expansion $\widetilde{\theta}$ (without torsion) reads
\begin{eqnarray} \label{eq:30}
\widetilde{\theta} &=& h^{\alpha \beta} \widetilde{B}_{\perp \alpha \beta} = h^{\alpha \beta} \widetilde{\nabla}_\alpha v_\beta \ .
\end{eqnarray}
With the help of Eq.~\eqref{eq:34}, the expansion $\theta$ can be written as
\begin{eqnarray} \label{eq:31}
\theta &=& h^{\alpha \beta} B_{\perp \alpha \beta} = h^{\alpha \beta} \left( \widetilde{\nabla}_\alpha v_\beta 
- {K_{\alpha \beta}}^\sigma v_\sigma
+ 2 S_{\rho \alpha \beta} v^\rho 
+ 2 S_{\rho \alpha \sigma} v^\rho v^\sigma v_\beta 
- {K_{\rho \beta}}^\sigma  v^\rho v_\sigma v_\alpha \right)  \ .
\end{eqnarray}
Using the antisymmetric properties of contorsion and torsion tensor one can easily see that the contraction of all terms in the parenthesis with projection metric does vanish except the first term. Therefore we get
\begin{eqnarray}
\theta &=& h^{\alpha \beta} \widetilde{\nabla}_\alpha v_\beta = \widetilde{\theta}
\end{eqnarray}

So we see that the expansion is the same with or without torsion~(this result confirms the one of~\cite{Luz2017}). The rate of change of $\theta$ w.r.t. proper time~($\tau$) along the timelike congruence reads
\begin{eqnarray} \label{eq:32}
\frac{D \theta}{d \tau} &=& v^\mu \nabla_\mu \theta = v^\mu \partial_\mu \theta = v^\mu \partial_\mu \widetilde{\theta} = v^\mu \widetilde{\nabla}_\mu \widetilde{\theta} = \frac{\widetilde{D} \widetilde{\theta}}{d \tau} \ .
\end{eqnarray}
$\frac{\widetilde{D} \widetilde{\theta}}{d \tau}$ is the Raychaudhuri equation for expansion without torsion,
so we see that Raychaudhuri equation for expansion for timelike congruence is unchanged
in the presence of torsion. 
So we get
\begin{eqnarray} \label{eq:33}
\frac{D \theta}{d \tau} = \frac{\widetilde{D} \widetilde{\theta}}{d \tau} =
- \widetilde{R}_{\gamma \rho} v^\rho v^\gamma 
- \left(\frac{1}{N-1} \widetilde{\theta}^2 + \widetilde{\sigma}_{\alpha \beta} \widetilde{\sigma}^{\alpha \beta} + \widetilde{\omega}_{\alpha \beta} \widetilde{\omega}^{\beta \alpha} \right)  \ ,
\end{eqnarray}
where  \hspace{0.1 cm} $\widetilde{•}$ \hspace{0.1 cm} quantities are calculated w.r.t. the Levi--Civita connection.
Let us stress that while this equation is identical to the one obtained without torsion, it will in general lead to a different phenomenology once the dynamics, i.e.~the field equation relating the Ricci tensor to the matter content, are used.  We shall discuss this in detail later on.
\subsection{Raychaudhuri equation for the congruence shear}
According to the definition of shear given in~\eqref{eq:22} and using~\eqref{eq:29},~\eqref{eq:30}, the expression of the shear without torsion reads
\begin{eqnarray} \label{eq:35}
\widetilde{\sigma}_{\alpha \beta} &=& \widetilde{\nabla}_{(\alpha} v_{\beta)}
- \frac{1}{N-1} h_{\alpha \beta} \widetilde{\nabla}_\alpha v^\alpha \ .
\end{eqnarray}
Using~\eqref{eq:22} and~\eqref{eq:34} the expression of shear reads
\begin{eqnarray} \label{eq:36}
\sigma_{\alpha \beta} &=& \widetilde{\sigma}_{\alpha \beta}
+ F_{\alpha \beta} \ ,
\end{eqnarray}
where
\begin{eqnarray} \label{eq:37}
F_{\alpha \beta} =
- {K_{(\alpha \beta)}}^\sigma v_\sigma
+ 2 S_{\rho (\alpha \beta)} v^\rho 
+ 2 S_{\rho (\alpha | \sigma} v^\rho v^\sigma v_{| \beta)} 
- {K_{\rho (\beta |}}^\sigma  v^\rho v_\sigma v_{| \alpha)} \ .
\end{eqnarray}
Now we calculate the Raychaudhuri equation governing the shear using~\eqref{eq:36}
\begin{eqnarray} \label{eq:38}
\frac{D \sigma_{\alpha \beta}}{d \tau} &=& v^\mu \nabla_\mu \sigma_{\alpha \beta} \nonumber \\
&=& v^\mu \widetilde{\nabla}_\mu \widetilde{\sigma}_{\alpha \beta}
+ v^\mu \widetilde{\nabla}_\mu F_{\alpha \beta}
- v^\mu {K_{\mu \alpha}}^\rho \widetilde{\sigma}_{\rho \beta}
- v^\mu {K_{\mu \alpha}}^\rho  F_{\rho \beta}
- v^\mu {K_{\mu \beta}}^\rho \widetilde{\sigma}_{\alpha \rho}
- v^\mu {K_{\mu \beta}}^\rho F_{\alpha \rho} \nonumber \\
&=& \frac{\widetilde{D} \widetilde{\sigma}_{\alpha \beta}}{d \tau}
+ v^\mu \widetilde{\nabla}_\mu F_{\alpha \beta}
- v^\mu {K_{\mu \alpha}}^\rho \widetilde{\sigma}_{\rho \beta}
- v^\mu {K_{\mu \alpha}}^\rho  F_{\rho \beta}
- v^\mu {K_{\mu \beta}}^\rho \widetilde{\sigma}_{\alpha \rho}
- v^\mu {K_{\mu \beta}}^\rho F_{\alpha \rho}
\end{eqnarray}
Substituting Raychaudhuri equation for shear~$\left(\frac{\widetilde{D} \widetilde{\sigma}_{\alpha \beta}}{d \tau}\right)$ we get
\begin{eqnarray} \label{eq:39}
\frac{D \sigma_{\alpha \beta}}{d \tau} &=& \Big[ - \frac{2}{N-1} \widetilde{\theta} \widetilde{\sigma}_{\alpha \beta}
- {\widetilde{\sigma}_\alpha}^{\,\,\,\gamma}  \widetilde{\sigma}_{\gamma \beta}
- {\widetilde{\omega}_\alpha}^{\,\,\,\gamma} \widetilde{\omega}_{\gamma \beta}
+ \frac{1}{N-1} h_{\alpha \beta} \left( \widetilde{\sigma}_{\gamma \rho} \widetilde{\sigma}^{\gamma \rho} 
- \widetilde{\omega}_{\gamma \rho} \widetilde{\omega}^{\gamma \rho} \right)
 - \widetilde{C}_{\alpha \gamma \beta \rho} v^\rho v^\gamma
+ \frac{1}{N-2} \widetilde{R}_{\alpha \beta}^{T} \nonumber \\
&& - \frac{1}{\left(N-1\right) \left(N-2\right)} h_{\alpha \beta} \left( \widetilde{R}_{\rho \gamma} h^{\rho \gamma} \right) \Big]
+ v^\mu \widetilde{\nabla}_\mu F_{\alpha \beta}
- v^\mu {K_{\mu \alpha}}^\rho \widetilde{\sigma}_{\rho \beta}
- v^\mu {K_{\mu \alpha}}^\rho  F_{\rho \beta}
- v^\mu {K_{\mu \beta}}^\rho \widetilde{\sigma}_{\alpha \rho} \nonumber \\
&& - v^\mu {K_{\mu \beta}}^\rho F_{\alpha \rho} \ ,
\end{eqnarray}
where $\widetilde{R}_{\alpha \beta}^{T} = h_{\alpha}^{\,\,\,\rho} h_{\beta}^{\,\,\,\gamma} \widetilde{R}_{\rho \gamma} $ and $\widetilde{C}_{\alpha \gamma \beta \rho}$ is the usual Weyl tensor. This is the Raychaudhuri equation for the shear in the presence of torsion, all the additional terms w.r.t the standard ones being grouped outside of the squared brackets.

\subsection{Raychaudhuri equation for the congruence vorticity}

Using~\eqref{eq:23} and~\eqref{eq:29}, the expression for vorticity without torsion reads
\begin{eqnarray} \label{eq:40}
\widetilde{\omega}_{\alpha \beta} &=& \widetilde{\nabla}_{[\alpha} v_{\beta]} \ .
\end{eqnarray}
Using~\eqref{eq:21} and~\eqref{eq:34} we get,
\begin{eqnarray} \label{eq:41}
\omega_{\alpha \beta} &=& \widetilde{\omega}_{\alpha \beta}
+ G_{\alpha \beta} \ ,
\end{eqnarray}
where,
\begin{eqnarray} \label{eq:42}
G_{\alpha \beta} &=& - {K_{[\alpha \beta]}}^\sigma v_\sigma
+ 2 S_{\rho [\alpha \beta]} v^\rho 
+ 2 S_{\rho [\alpha| \sigma} v^\rho v^\sigma v_{|\beta]} 
- {K_{\rho [\beta|}}^\sigma  v^\rho v_\sigma v_{|\alpha]}
\end{eqnarray}
We get the Raychaudhuri equation for vorticity using~\eqref{eq:41}
\begin{eqnarray} \label{eq:43}
\frac{D \omega_{\alpha \beta}}{d \tau} &=& v^\mu \nabla_\mu \omega_{\alpha \beta} \nonumber \\
&=& v^\mu \widetilde{\nabla}_\mu \widetilde{\omega}_{\alpha \beta}
+ v^\mu \widetilde{\nabla}_\mu G_{\alpha \beta}
- v^\mu {K_{\mu \alpha}}^\rho \widetilde{\omega}_{\rho \beta}
- v^\mu {K_{\mu \alpha}}^\rho  G_{\rho \beta}
- v^\mu {K_{\mu \beta}}^\rho \widetilde{\omega}_{\alpha \rho}
- v^\mu {K_{\mu \beta}}^\rho G_{\alpha \rho} \nonumber \\
&=& \frac{\widetilde{D} \widetilde{\omega}_{\alpha \beta}}{d \tau}
+ v^\mu \widetilde{\nabla}_\mu G_{\alpha \beta}
- v^\mu {K_{\mu \alpha}}^\rho \widetilde{\omega}_{\rho \beta}
- v^\mu {K_{\mu \alpha}}^\rho  G_{\rho \beta}
- v^\mu {K_{\mu \beta}}^\rho \widetilde{\omega}_{\alpha \rho}
- v^\mu {K_{\mu \beta}}^\rho G_{\alpha \rho}
\end{eqnarray}
Substituting the Raychaudhuri equation without torsion~$\left( \frac{\widetilde{D} \widetilde{\omega}_{\alpha \beta}}{d \tau} \right)$ we get
\begin{eqnarray} \label{eq:44}
\frac{D \omega_{\alpha \beta}}{d \tau} &=& \big[- \frac{2}{N-1} \widetilde{\theta} \widetilde{\omega}_{\alpha\beta}
- \widetilde{\sigma}_{\alpha\gamma} {\widetilde{\omega}^\gamma}_\beta
- \widetilde{\omega}_{\alpha\gamma} {\widetilde{\sigma}^\gamma}_\beta \big]
+ v^\mu \widetilde{\nabla}_\mu G_{\alpha \beta}
- v^\mu {K_{\mu \alpha}}^\rho \widetilde{\omega}_{\rho \beta}
- v^\mu {K_{\mu \alpha}}^\rho  G_{\rho \beta}
- v^\mu {K_{\mu \beta}}^\rho \widetilde{\omega}_{\alpha \rho} \nonumber \\
&& - v^\mu {K_{\mu \beta}}^\rho G_{\alpha \rho} \ ,
\end{eqnarray}
where in squared bracket we grouped the analogous equation for the torsion-free case.

\section{Raychaudhuri equation with torsion for null congruences}\label{null}

Let us consider an auxiliary null vector field $\epsilon^\alpha$ , such that
\begin{eqnarray}
k^\alpha \epsilon_\alpha &=& -1 \ , \label{eq:45} \\
\epsilon_\alpha \epsilon^\alpha &=& 0 \label{eq:46} \ .
\end{eqnarray}
The projector operator is defined as
\begin{eqnarray} \label{eq:47}
\widetilde{h}_{\alpha \beta} = g_{\alpha \beta} + k_\alpha \epsilon_\beta + \epsilon_\alpha k_\beta
\end{eqnarray}
satisfying
\begin{eqnarray} 
\widetilde{h}_{\alpha \beta} k^\alpha = \widetilde{h}_{\alpha \beta} \epsilon^\alpha = 0 \ , \label{eq:48} \\
\widetilde{h}^{\,\,\,\,\sigma}_\alpha \widetilde{h}_{\sigma \beta} = \widetilde{h}_{\alpha \beta} \ , \label{eq:49} \\
\widetilde{h}^{\,\,\,\,\alpha}_\alpha = N-2 \label{eq:50} \ .
\end{eqnarray}
Photon follows geodesics determined by Levi--Civita connection,
\begin{eqnarray} \label{eq:51}
k^\alpha \widetilde{\nabla}_\alpha k^\beta = 0 \ .
\end{eqnarray}
The perpendicular component of $B_{\alpha \beta}$ can be calculated by
%
\begin{eqnarray} \label{eq:52}
B_{\perp \alpha \beta} &=& h^{\,\,\,\gamma}_\alpha h^{\,\,\,\sigma}_\beta B_{\gamma \sigma} \nonumber  \\
&=& \widetilde{\nabla}_\alpha k_\beta
- {K_{\alpha \beta}}^{\rho} k_\rho
- 2 S_{\alpha\gamma\beta} k^\gamma
- 2 S_{\alpha \gamma \sigma} k^\gamma k^\sigma \epsilon_\beta
+ \epsilon^\sigma k_\beta \widetilde{\nabla}_\alpha k_\sigma
- \epsilon^\sigma k_\beta {K_{\alpha \sigma}}^{\rho} k_\rho
- 2 S_{\alpha \gamma \sigma} k^\gamma k_\beta \epsilon^\sigma
+ \epsilon^\gamma k_\alpha \widetilde{\nabla}_\gamma k_\beta \nonumber \\
&& - \epsilon^\gamma k_\alpha {K_{\gamma \beta}}^{\rho} k_\rho
+ 2 S_{\rho \gamma \beta} k^\rho k_\alpha \epsilon^\gamma
+ 2 S_{\rho \gamma \sigma} k^\rho k_\alpha k^\sigma \epsilon^\gamma \epsilon_\beta
+ \epsilon^\gamma k_\alpha \epsilon^\sigma k_\beta \widetilde{\nabla}_\gamma k_\sigma
- \epsilon^\gamma k_\alpha \epsilon^\sigma k_\beta {K_{\gamma \sigma}}^{\rho} k_\rho
+ 2 S_{\rho \gamma \sigma} k^\rho k_\alpha k_\beta \epsilon^\gamma \epsilon^\sigma \nonumber \\
&& + 2 S_{\beta \gamma  \sigma} k^\sigma k^\gamma \epsilon_\alpha
+ 2 k_\beta \epsilon_\alpha \epsilon^\sigma S_{\sigma \gamma \rho} k^\rho k^\gamma \ .
\end{eqnarray}
The perpendicular component of $B_{\alpha \beta}$ without
presence of torsion can be written
\begin{eqnarray} \label{eq:53}
\widetilde{B}_{\perp \alpha \beta} &=& \widetilde{\nabla}_\alpha k_\beta
+ \epsilon^\sigma k_\beta \widetilde{\nabla}_\alpha k_\sigma
+ \epsilon^\gamma k_\alpha \widetilde{\nabla}_\gamma k_\beta
+ \epsilon^\gamma k_\alpha \epsilon^\sigma k_\beta \widetilde{\nabla}_\gamma k_\sigma \ .
\end{eqnarray}

\subsection{Raychaudhuri equation for the congruence expansion}
Following the definition of expansion given in~\eqref{eq:21} and using~\eqref{eq:53} we get
\begin{eqnarray} \label{eq:54}
\widetilde{\theta} &=& h^{\alpha \beta} \widetilde{B}_{\perp \alpha \beta} = h^{\alpha \beta} \widetilde{\nabla}_\alpha k_\beta \ .
\end{eqnarray}
Similarly like timelike congruence we can write using~\eqref{eq:21} and~\eqref{eq:52}
\begin{eqnarray} \label{eq:55}
\theta &=& h^{\alpha \beta} B_{\perp \alpha \beta} =  h^{\alpha \beta} \widetilde{\nabla}_\alpha k_\beta = \widetilde{\theta} \ .
\end{eqnarray}
So even in the case of null congruence expansion does not get affected due to the presence of torsion. Evolution of expansion along the null congruence reads 
\begin{eqnarray} \label{eq:56}
\frac{D \theta}{d \lambda} &=& k^\mu \nabla_\mu \theta = k^\mu \partial_\mu \theta = k^\mu \partial_\mu \widetilde{\theta} = k^\mu \widetilde{\nabla}_\mu \widetilde{\theta} = \frac{\widetilde{D} \widetilde{\theta}}{d \lambda} \ .
\end{eqnarray}
Substituting Raychaudhuri equation for expansion $\left( \frac{\widetilde{D} \widetilde{\theta}}{d \lambda} \right)$~\cite{Luz2017} we get
\begin{eqnarray} \label{eq:57}
\frac{D \theta}{d \lambda} = \frac{\widetilde{D} \widetilde{\theta}}{d \lambda} = - \widetilde{R}_{\eta \rho} k^\eta k^\rho
-\left(\frac{1}{N-2} \widetilde{\theta}^2 + \widetilde{\sigma}_{\alpha \beta} \widetilde{\sigma}^{\alpha \beta} + \widetilde{\omega}_{\alpha \beta} \widetilde{\omega}^{\beta \alpha} \right) \ .
\end{eqnarray}
Eq.~\eqref{eq:57} is the Raychaudhuri equation for expansion in the case of null congruence.
As in the case of the timelike analogue expression, it is worth stressing that while the above equation is identical to the one obtained without torsion, it will in general lead to a different phenomenology once the dynamics, i.e.~the field equation relating the Ricci tensor to the matter content, are used.

\subsection{Raychaudhuri equation for the congruence shear}

Using~\eqref{eq:22} and~\eqref{eq:53} we get expression of
shear without torsion reads
\begin{eqnarray} \label{eq:58}
\widetilde{\sigma}_{\alpha \beta} = \widetilde{\nabla}_{(\alpha} k_{\beta)}
+ \epsilon^\sigma k_{(\beta |} \widetilde{\nabla}_{| \alpha)} k_\sigma
+ \epsilon^\gamma k_{(\alpha |} \widetilde{\nabla}_\gamma k_{| \beta)}
+ \epsilon^\gamma k_{(\alpha |} \epsilon^\sigma k_{| \beta)} \widetilde{\nabla}_\gamma k_\sigma
- \frac{1}{N-2} h_{\alpha \beta} \widetilde{\nabla}_\alpha k^\alpha \ .
\end{eqnarray}
Using~\eqref{eq:22} and~\eqref{eq:52} we get
\begin{eqnarray} \label{eq:59}
\sigma_{\alpha \beta} = \widetilde{\sigma}_{\alpha \beta}
+ H_{\alpha \beta} \ ,
\end{eqnarray}
where,
\begin{eqnarray} \label{eq:60}
H_{\alpha \beta} &=& - {K_{(\alpha \beta)}}^{\rho} k_\rho
- 2 S_{(\alpha | \gamma | \beta)} k^\gamma
- 2 S_{(\alpha | \gamma \sigma} k^\gamma k^\sigma \epsilon_{| \beta)}
- \epsilon^\sigma k_{(\beta |} {K_{| \alpha) \sigma}}^{\rho} k_\rho
- 2 S_{(\alpha | \gamma \sigma} k^\gamma k_{| \beta)} \epsilon^\sigma \nonumber \\
&& - \epsilon^\gamma k_{(\alpha |} {K_{\gamma | \beta )}}^{\rho} k_\rho 
+ 2 S_{\rho \gamma (\beta |} k^\rho k_{| \alpha)} \epsilon^\gamma
+ 2 S_{\rho \gamma \sigma} k^\rho k_{(\alpha |} k^\sigma \epsilon^\gamma \epsilon_{| \beta)}
- \epsilon^\gamma k_{(\alpha |} \epsilon^\sigma k_{| \beta)} {K_{\gamma \sigma}}^{\rho} k_\rho
+ 2 S_{\rho \gamma \sigma} k^\rho k_{(\alpha |} k_{| \beta)} \epsilon^\gamma \epsilon^\sigma \nonumber \\
&& + 2 S_{( \beta | \gamma  \sigma} k^\sigma k^\gamma \epsilon_{| \alpha)}
+ 2 k_{(\beta |} \epsilon_{| \alpha)} \epsilon^\sigma S_{\sigma \gamma \rho} k^\rho k^\gamma
\end{eqnarray}
The evolution of shear along null congruence can be written as
\begin{eqnarray} \label{eq:61}
\frac{D \sigma_{\alpha \beta}}{d \lambda} &=& k^\mu \nabla_\mu \sigma_{\alpha \beta} \nonumber \\
&=& k^\mu \widetilde{\nabla}_\mu \widetilde{\sigma}_{\alpha \beta}
+ k^\mu \widetilde{\nabla}_\mu H_{\alpha \beta}
- k^\mu {K_{\mu \alpha}}^\rho \widetilde{\sigma}_{\rho \beta}
- k^\mu {K_{\mu \alpha}}^\rho  H_{\rho \beta}
- k^\mu {K_{\mu \beta}}^\rho \widetilde{\sigma}_{\alpha \rho}
- k^\mu {K_{\mu \beta}}^\rho H_{\alpha \rho} \nonumber \\
&=& \frac{\widetilde{D} \widetilde{\sigma}_{\alpha \beta}}{d \lambda}
+ k^\mu \widetilde{\nabla}_\mu H_{\alpha \beta}
- k^\mu {K_{\mu \alpha}}^\rho \widetilde{\sigma}_{\rho \beta}
- k^\mu {K_{\mu \alpha}}^\rho  H_{\rho \beta}
- k^\mu {K_{\mu \beta}}^\rho \widetilde{\sigma}_{\alpha \rho}
- k^\mu {K_{\mu \beta}}^\rho H_{\alpha \rho} 
\end{eqnarray}
Substituting the Raychaudhuri equation for shear without torsion~$\left(\frac{\widetilde{D} \widetilde{\sigma}_{\alpha \beta}}{d \lambda}\right)$ we get
\begin{eqnarray} \label{eq:62}
\frac{D \sigma_{\alpha \beta}}{d \lambda} &=& \Bigg[ - \frac{2}{N-2} \widetilde{\theta} \widetilde{\sigma}_{\alpha \beta}
- {\widetilde{\sigma}_\alpha}^{\,\,\,\gamma}  \widetilde{\sigma}_{\gamma \beta}
- {\widetilde{\omega}_\alpha}^{\,\,\,\gamma} \widetilde{\omega}_{\gamma \beta}
+ \frac{1}{N-2} h_{\alpha \beta} \left( \widetilde{\sigma}_{\gamma \rho} \widetilde{\sigma}^{\gamma \rho} 
- \widetilde{\omega}_{\gamma \rho} \widetilde{\omega}^{\gamma \rho} \right)
- \widetilde{C}_{\alpha \gamma \beta \rho} k^\rho k^\gamma
+ \frac{1}{N-2} \widetilde{R}_{\alpha \beta}^{T} \nonumber \\
&& + \frac{1}{\left(N-2\right)^2} h_{\alpha \beta} \left( \widetilde{R}_{\rho \gamma} h^{\rho \gamma} \right) \Bigg]
+ k^\mu \widetilde{\nabla}_\mu H_{\alpha \beta}
- k^\mu {K_{\mu \alpha}}^\rho \widetilde{\sigma}_{\rho \beta}
- k^\mu {K_{\mu \alpha}}^\rho  H_{\rho \beta}
- k^\mu {K_{\mu \beta}}^\rho \widetilde{\sigma}_{\alpha \rho}
- k^\mu {K_{\mu \beta}}^\rho H_{\alpha \rho} \ .
\end{eqnarray}
The above equation is the Raychaudhuri equation for shear in the presence of torsion. Again the expression in squared brackets is the usual one for the torsion-free case.

\subsection{Raychaudhuri equation for the congruence vorticity}

Following the definition of vorticity given in~\eqref{eq:23} and~\eqref{eq:52} we get
\begin{eqnarray} \label{eq:63}
\widetilde{\omega}_{\alpha \beta} = \widetilde{\nabla}_{[\alpha} k_{\beta]}
+ \epsilon^\sigma k_{[\beta} \widetilde{\nabla}_{\alpha]} k_\sigma
+ \epsilon^\gamma k_{[\alpha|} \widetilde{\nabla}_\gamma k_{|\beta]} \ .
\end{eqnarray}
Using~\eqref{eq:23} and~\eqref{eq:53} the expression of vorticity can be written as
\begin{eqnarray} \label{eq:64}
\omega_{\alpha \beta} = \widetilde{\omega}_{\alpha \beta}
+ W_{\alpha \beta} \ ,
\end{eqnarray}
where,
\begin{eqnarray} \label{eq:65}
W_{\alpha \beta} &=& - {K_{[\alpha \beta]}}^{\rho} k_\rho
- 2 S_{[\alpha | \gamma | \beta]} k^\gamma
- 2 S_{[\alpha | \gamma \sigma} k^\gamma k^\sigma \epsilon_{|\beta]}
- \epsilon^\sigma k_{[\beta} {K_{\alpha ] \sigma}}^{\rho} k_\rho
- 2 S_{[\alpha | \gamma \sigma} k^\gamma k_{| \beta]} \epsilon^\sigma
- \epsilon^\gamma k_{[\alpha|} {K_{\gamma | \beta]}}^{\rho} k_\rho
+ 2 S_{\rho \gamma [\beta | } k^\rho k_{|\alpha]} \epsilon^\gamma \nonumber \\
&& + 2 S_{\rho \gamma \sigma} k^\rho k_{[\alpha|} k^\sigma \epsilon^\gamma \epsilon_{|\beta]} 
+ 2 S_{[\beta | \gamma  \sigma} k^\sigma k^\gamma \epsilon_{|\alpha]}
+ 2 k_{[\beta} \epsilon_{\alpha]} \epsilon^\sigma S_{\sigma \gamma \rho} k^\rho k^\gamma \ .
\end{eqnarray}
The evolution of vorticity along null congruence is given by
\begin{eqnarray} \label{eq:66}
\frac{D \omega_{\alpha \beta}}{d \lambda} &=& k^\mu \nabla_\mu \omega_{\alpha \beta} \nonumber \\
&=& k^\mu \widetilde{\nabla}_\mu \widetilde{\omega}_{\alpha \beta}
+ k^\mu \widetilde{\nabla}_\mu W_{\alpha \beta}
- k^\mu {K_{\mu \alpha}}^\rho \widetilde{\omega}_{\rho \beta}
- k^\mu {K_{\mu \alpha}}^\rho  W_{\rho \beta}
- k^\mu {K_{\mu \beta}}^\rho \widetilde{\omega}_{\alpha \rho}
- k^\mu {K_{\mu \beta}}^\rho W_{\alpha \rho} \nonumber \\
&=& \frac{\widetilde{D} \widetilde{\omega}_{\alpha \beta}}{d \lambda}
+ k^\mu \widetilde{\nabla}_\mu W_{\alpha \beta}
- k^\mu {K_{\mu \alpha}}^\rho \widetilde{\omega}_{\rho \beta}
- k^\mu {K_{\mu \alpha}}^\rho  W_{\rho \beta}
- k^\mu {K_{\mu \beta}}^\rho \widetilde{\omega}_{\alpha \rho}
- k^\mu {K_{\mu \beta}}^\rho W_{\alpha \rho} \ .
\end{eqnarray}
Substituting the Raychaudhuri equation for vorticity without torsion~$\left( \frac{\widetilde{D} \widetilde{\omega}_{\alpha \beta}}{d \lambda} \right)$ we get
\begin{eqnarray}  \label{eq:67}
\frac{D \omega_{\alpha \beta}}{d \lambda} &=& \Big[- \frac{2}{N-2} \widetilde{\theta} \widetilde{\omega}_{\alpha\beta}
- \widetilde{\sigma}_{\alpha\gamma} {\widetilde{\omega}^{\,\,\,\gamma}}_\beta
- \widetilde{\omega}_{\alpha\gamma} {\widetilde{\sigma}^{\,\,\,\gamma}}_\beta \Big]
+ k^\mu \widetilde{\nabla}_\mu W_{\alpha \beta}
- k^\mu {K_{\mu \alpha}}^\rho \widetilde{\omega}_{\rho \beta}
- k^\mu {K_{\mu \alpha}}^\rho  W_{\rho \beta}
- k^\mu {K_{\mu \beta}}^\rho \widetilde{\omega}_{\alpha \rho} \nonumber \\
&& - k^\mu {K_{\mu \beta}}^\rho W_{\alpha \rho} \ .
\end{eqnarray}
Above equation is the Raychaudhuri equation for vorticity in the presence of torsion with the torsion-free expression separated in squared brackets~\footnote{To the best of our knowledge the $N$-dimensional Raychaudhuri equation for vorticity and shear in the case of both timelike and null congruences are original at least in the form we cast here.}.

\section{Einstein--Cartan--Sciama--Kibble Field Equations}\label{sec:ECSK}

We have already stressed that the Raychaudhuri equations are exquisitely geometrical in nature, in the sense that they do not depend on the specific field equations of the gravitational theory describing the dynamics of the metric and torsion. However, when describing the outcome of physical phenomena like a gravitational collapse one needs to connect these equations with the matter-energy content, and hence the gravitational field equations must be supplemented. In particular, for our purposes we shall need to consider the Raychaudhuri equation for the expansions of null congruence (as in Penrose's theorem) and remember, as anticipated, that in a theory with torsion, such as ECT, the Ricci tensor $\widetilde{R}_{\gamma \rho}$ will not be the same as in Einstein gravity. 

Among the possible theories of gravity with torsion we consider ECT for its simplicity and naturalness as an extension of GR in this setting. Indeed, in ECT torsion does not propagate in vacuum, but it is generated dynamically in the presence of matter with spin. The generalization of Einstein's field equations for particle with intrinsic spin is known as Einstein--Cartan--Sciama--Kibble equations~\cite{Sciama1962,Kibble1961} read
\begin{eqnarray} 
G_{\mu \nu} = k \Sigma_{\mu \nu} \label{eq:68} \ , \\
T_{\mu \nu}^{\,\,\,\,\,\,\gamma} = k \tau_{\mu \nu}^{\,\,\,\,\,\,\gamma} \ ,\label{eq:69}
\end{eqnarray}
where,
$k = 8 \pi G/c^4$, $G_{\mu \nu}$ is Einstein tensor, and
\begin{eqnarray} \label{eq:70}
T_{\mu \nu}^{\,\,\,\,\,\,\gamma} = S_{\mu \nu}^{\,\,\,\,\,\,\gamma} + \delta_\mu^{\,\,\,\gamma} S_{\nu \zeta}^{\,\,\,\,\,\,\zeta} - \delta_\nu^{\,\,\,\gamma} S_{\mu \zeta}^{\,\,\,\,\,\,\zeta} \ ,
\end{eqnarray}
is the modified torsion tensor. $\Sigma_{\mu \nu}$ is the modified stress-energy tensor (SET) given by~\cite{Hehl1976}
\begin{eqnarray} \label{eq:72}
\Sigma_{\mu \nu} = t_{\mu \nu} + \nabla_\xi \left( \tau_{\mu \nu}^{\,\,\,\,\,\,\xi} - \tau_{\nu \,\,\,\mu}^{\,\,\,\xi} + \tau^\xi_{\,\,\,\mu \nu} \right) \ ,
\end{eqnarray}
where
$\nabla_\xi$ is the covariant derivative with the presence of torsion and $\widetilde{\nabla}_\xi$ is the covariant derivative without torsion related by $\nabla_\xi \equiv \widetilde{\nabla}_\xi + 2 S_{\xi \nu}^{\,\,\,\,\,\,\nu} $ and
\begin{eqnarray}
t^{\mu \nu} &\equiv& \frac{2}{e}\frac{\delta \mathcal{L}}{\delta g_{\mu \nu}} \quad \textrm{(stress energy tensor)} \ , \label{eq:73} \\
\tau_k^{\,\,\,\nu \mu} &\equiv& \frac{1}{e} \frac{\delta \mathcal{L}}{\delta K_{\mu \nu}^{\,\,\,\,\,\,k}} \quad \textrm{(spin angular momentum tensor)}\ , \label{eq:74}
\end{eqnarray}
while
$e = \sqrt{\mathrm{det}(g_{\mu \nu})}$,
$\mathcal{L}$ is the matter Lagrangian and $ K_{\mu \nu}^{\,\,\,\,\,\,k}$ is the contorsion tensor given by Eq.~\eqref{eq:6}. Eq.~\eqref{eq:69} can be solved to give~\cite{Trautman2006}
\begin{eqnarray} \label{eq:71}
S_{\mu \nu}^{\,\,\,\,\,\,\xi} = k \left( \tau_{\mu \nu}^{\,\,\,\,\,\,\xi}
+ \frac{1}{2} \delta_\mu^{\,\,\,\xi} \tau_{\nu \sigma}^{\,\,\,\,\,\,\sigma}
+ \frac{1}{2} \delta_\nu^{\,\,\,\xi} \tau_{\sigma \mu}^{\,\,\,\,\,\,\sigma} \right) \ .
\end{eqnarray}
\\

The Ricci tensor can be written from Eq.~\eqref{eq:68} as
\begin{eqnarray} \label{eq:75}
R_{\mu \nu} &=& k \Sigma_{\mu \nu} - \frac{1}{2} k g_{\mu \nu} \Sigma \ ,
\end{eqnarray}
where $\Sigma = g^{\mu \nu} \Sigma_{\mu \nu}$. By expanding the l.h.s. of Eq.~\eqref{eq:75} and rearranging, the Ricci tensor w.r.t. Levi--Civita connection can be written as
\begin{eqnarray} \label{eq:76}
\widetilde{R}_{\mu \nu} &=& k \Sigma_{\mu \nu} - \frac{1}{2} k g_{\mu \nu} \Sigma
- \left( \partial_\sigma {K_{\mu \nu}}^\sigma - \partial_\mu {K_{\sigma \nu}}^\sigma + {K_{\mu \nu}}^\epsilon {K_{\delta \epsilon}}^\delta - {K_{\delta \nu}}^\epsilon {K_{\mu \epsilon}}^\delta \right) \nonumber \\
&& - \left( \Gamma^{\epsilon}_{\,\,\,\mu \nu} {K_{\delta \epsilon}}^\delta + {K_{\mu \nu}}^\epsilon \Gamma^{\delta}_{\,\,\,\delta \epsilon} - \Gamma^{\epsilon}_{\,\,\,\delta \nu} {K_{\mu \epsilon}}^\delta - {K_{\delta \nu}}^\epsilon \Gamma^{\delta}_{\,\,\,\mu \epsilon} \right) \ .
\end{eqnarray}
In Eq.~\eqref{eq:76}, we see that the l.h.s is symmetric being the Ricci tensor of the Levi--Civita connection. So the r.h.s of Eq.~\eqref{eq:76} should be symmetric as well. Let us then write the terms or the r.h.s of the above equation as combination of symmetric and antisymmetric tensors as
\begin{eqnarray} \label{eq:77}
&& \widetilde{R}_{\mu \nu} = \frac{1}{2} k \left( \Sigma_{\mu \nu} + \Sigma_{\nu \mu} \right) + \frac{1}{2} k \left( \Sigma_{\mu \nu} - \Sigma_{\nu \mu} \right)
- \frac{1}{2} k g_{\mu \nu} \Sigma \nonumber \\
&& - \frac{1}{2} \left[ \left( \partial_\sigma {K_{\mu \nu}}^\sigma - \partial_\mu {K_{\sigma \nu}}^\sigma + {K_{\mu \nu}}^\epsilon {K_{\delta \epsilon}}^\delta - {K_{\delta \nu}}^\epsilon {K_{\mu \epsilon}}^\delta \right)
+ \left( \partial_\sigma {K_{\nu\mu}}^\sigma - \partial_\nu {K_{\sigma \mu}}^\sigma + {K_{\nu\mu}}^\epsilon {K_{\delta \epsilon}}^\delta - {K_{\delta \mu}}^\epsilon {K_{\nu \epsilon}}^\delta \right) \right] \nonumber \\
&& - \frac{1}{2} \left[ \left( \partial_\sigma {K_{\mu \nu}}^\sigma - \partial_\mu {K_{\sigma \nu}}^\sigma + {K_{\mu \nu}}^\epsilon {K_{\delta \epsilon}}^\delta - {K_{\delta \nu}}^\epsilon {K_{\mu \epsilon}}^\delta \right)
- \left( \partial_\sigma {K_{\nu\mu}}^\sigma - \partial_\nu {K_{\sigma \mu}}^\sigma + {K_{\nu\mu}}^\epsilon {K_{\delta \epsilon}}^\delta - {K_{\delta \mu}}^\epsilon {K_{\nu \epsilon}}^\delta \right) \right] \nonumber \\
&& - \frac{1}{2} \left[ \left( \Gamma^{\epsilon}_{\,\,\,\mu \nu} {K_{\delta \epsilon}}^\delta + {K_{\mu \nu}}^\epsilon \Gamma^{\delta}_{\,\,\,\delta \epsilon} - \Gamma^{\epsilon}_{\,\,\,\delta \nu} {K_{\mu \epsilon}}^\delta - {K_{\delta \nu}}^\epsilon \Gamma^{\delta}_{\,\,\,\mu \epsilon} \right)
+ \left( \Gamma^{\epsilon}_{\,\,\,\nu\mu} {K_{\delta \epsilon}}^\delta + {K_{\nu\mu}}^\epsilon \Gamma^{\delta}_{\,\,\,\delta \epsilon} - \Gamma^{\epsilon}_{\,\,\,\delta \mu} {K_{\nu \epsilon}}^\delta - {K_{\delta \mu}}^\epsilon \Gamma^{\delta}_{\,\,\,\nu \epsilon} \right) \right] \nonumber \\
&& - \frac{1}{2} \left[ \left( \Gamma^{\epsilon}_{\,\,\,\mu \nu} {K_{\delta \epsilon}}^\delta + {K_{\mu \nu}}^\epsilon \Gamma^{\delta}_{\,\,\,\delta \epsilon} - \Gamma^{\epsilon}_{\,\,\,\delta \nu} {K_{\mu \epsilon}}^\delta - {K_{\delta \nu}}^\epsilon \Gamma^{\delta}_{\,\,\,\mu \epsilon} \right)
- \left( \Gamma^{\epsilon}_{\,\,\,\nu\mu} {K_{\delta \epsilon}}^\delta + {K_{\nu\mu}}^\epsilon \Gamma^{\delta}_{\,\,\,\delta \epsilon} - \Gamma^{\epsilon}_{\,\,\,\delta \mu} {K_{\nu \epsilon}}^\delta - {K_{\delta \mu}}^\epsilon \Gamma^{\delta}_{\,\,\,\nu \epsilon} \right) \right] \ .
\end{eqnarray}
Rewriting the r.h.s of the Eq.~\eqref{eq:77} to be symmetric implies the condition
\begin{eqnarray} \label{eq:78}
&& k \left( \Sigma_{\mu \nu} - \Sigma_{\nu \mu} \right) \nonumber \\
&& = \left[ \left( \partial_\sigma {K_{\mu \nu}}^\sigma - \partial_\mu {K_{\sigma \nu}}^\sigma + {K_{\mu \nu}}^\epsilon {K_{\delta \epsilon}}^\delta - {K_{\delta \nu}}^\epsilon {K_{\mu \epsilon}}^\delta \right)
- \left( \partial_\sigma {K_{\nu\mu}}^\sigma - \partial_\nu {K_{\sigma \mu}}^\sigma + {K_{\nu\mu}}^\epsilon {K_{\delta \epsilon}}^\delta - {K_{\delta \mu}}^\epsilon {K_{\nu \epsilon}}^\delta \right) \right] \nonumber \\
&& + \left[ \left( \Gamma^{\epsilon}_{\,\,\,\mu \nu} {K_{\delta \epsilon}}^\delta + {K_{\mu \nu}}^\epsilon \Gamma^{\delta}_{\,\,\,\delta \epsilon} - \Gamma^{\epsilon}_{\,\,\,\delta \nu} {K_{\mu \epsilon}}^\delta - {K_{\delta \nu}}^\epsilon \Gamma^{\delta}_{\,\,\,\mu \epsilon} \right)
- \left( \Gamma^{\epsilon}_{\,\,\,\nu\mu} {K_{\delta \epsilon}}^\delta + {K_{\nu\mu}}^\epsilon \Gamma^{\delta}_{\,\,\,\delta \epsilon} - \Gamma^{\epsilon}_{\,\,\,\delta \mu} {K_{\nu \epsilon}}^\delta - {K_{\delta \mu}}^\epsilon \Gamma^{\delta}_{\,\,\,\nu \epsilon} \right) \right] \ .
\end{eqnarray}
Eq.~\eqref{eq:78} is indeed the generalisation of special relativistic conservation law of total angular momentum in a Riemann-Cartan geometry~\cite{Obukhov1987}.
Imposing this condition~\eqref{eq:78} on Eq.~\eqref{eq:77} we get
\begin{eqnarray} \label{eq:79}
&& \widetilde{R}_{\mu \nu} = \frac{1}{2} k \left( \Sigma_{\mu \nu} + \Sigma_{\nu \mu} \right)
- \frac{1}{2} k g_{\mu \nu} \Sigma \nonumber \\
&& - \frac{1}{2} \left[ \left( \partial_\sigma {K_{\mu \nu}}^\sigma - \partial_\mu {K_{\sigma \nu}}^\sigma + {K_{\mu \nu}}^\epsilon {K_{\delta \epsilon}}^\delta - {K_{\delta \nu}}^\epsilon {K_{\mu \epsilon}}^\delta \right)
+ \left( \partial_\sigma {K_{\nu\mu}}^\sigma - \partial_\nu {K_{\sigma \mu}}^\sigma + {K_{\nu\mu}}^\epsilon {K_{\delta \epsilon}}^\delta - {K_{\delta \mu}}^\epsilon {K_{\nu \epsilon}}^\delta \right) \right] \nonumber \\
&& - \frac{1}{2} \left[ \left( \Gamma^{\epsilon}_{\,\,\,\mu \nu} {K_{\delta \epsilon}}^\delta + {K_{\mu \nu}}^\epsilon \Gamma^{\delta}_{\,\,\,\delta \epsilon} - \Gamma^{\epsilon}_{\,\,\,\delta \nu} {K_{\mu \epsilon}}^\delta - {K_{\delta \nu}}^\epsilon \Gamma^{\delta}_{\,\,\,\mu \epsilon} \right)
+ \left( \Gamma^{\epsilon}_{\,\,\,\nu\mu} {K_{\delta \epsilon}}^\delta + {K_{\nu\mu}}^\epsilon \Gamma^{\delta}_{\,\,\,\delta \epsilon} - \Gamma^{\epsilon}_{\,\,\,\delta \mu} {K_{\nu \epsilon}}^\delta - {K_{\delta \mu}}^\epsilon \Gamma^{\delta}_{\,\,\,\nu \epsilon} \right) \right] \ .
\end{eqnarray}
Now the Einstein tensor of Levi--Civita connection takes the form
\begin{eqnarray} \label{eq:80}
\widetilde{G}_{\mu \nu} = \widetilde{R}_{\mu \nu} - \frac{1}{2} g_{\mu \nu} g^{\alpha \beta}\widetilde{R}_{\alpha \beta} \ .
\end{eqnarray}
So using~\eqref{eq:79} we can rewrite the field equations for ECT in a form which closely resemble the GR one
\begin{eqnarray} \label{eq:81}
&& \widetilde{G}_{\mu \nu} = k \Sigma_{\mu \nu}^{\textrm{eff}} \ ,
\end{eqnarray}
where,
\begin{eqnarray} \label{eq:82}
&& \Sigma_{\mu \nu}^{\textrm{eff}} = 
 \frac{1}{2} \left( \Sigma_{\mu \nu} + \Sigma_{\nu \mu} \right)
+ \frac{1}{2} g_{\mu \nu} \Sigma
- \frac{1}{4} g_{\mu \nu} g^{\alpha \beta} \left( \Sigma_{\alpha \beta} + \Sigma_{\beta \alpha} \right) \nonumber \\
&& - \frac{1}{2k} \left[ \left( \partial_\sigma {K_{\mu \nu}}^\sigma - \partial_\mu {K_{\sigma \nu}}^\sigma + {K_{\mu \nu}}^\epsilon {K_{\delta \epsilon}}^\delta - {K_{\delta \nu}}^\epsilon {K_{\mu \epsilon}}^\delta \right)
+ \left( \partial_\sigma {K_{\nu\mu}}^\sigma - \partial_\nu {K_{\sigma \mu}}^\sigma + {K_{\nu\mu}}^\epsilon {K_{\delta \epsilon}}^\delta - {K_{\delta \mu}}^\epsilon {K_{\nu \epsilon}}^\delta \right) \right] \nonumber \\
&& - \frac{1}{2k} \left[ \left( \Gamma^{\epsilon}_{\,\,\,\mu \nu} {K_{\delta \epsilon}}^\delta + {K_{\mu \nu}}^\epsilon \Gamma^{\delta}_{\,\,\,\delta \epsilon} - \Gamma^{\epsilon}_{\,\,\,\delta \nu} {K_{\mu \epsilon}}^\delta - {K_{\delta \nu}}^\epsilon \Gamma^{\delta}_{\,\,\,\mu \epsilon} \right)
+ \left( \Gamma^{\epsilon}_{\,\,\,\nu\mu} {K_{\delta \epsilon}}^\delta + {K_{\nu\mu}}^\epsilon \Gamma^{\delta}_{\,\,\,\delta \epsilon} - \Gamma^{\epsilon}_{\,\,\,\delta \mu} {K_{\nu \epsilon}}^\delta - {K_{\delta \mu}}^\epsilon \Gamma^{\delta}_{\,\,\,\nu \epsilon} \right) \right] \nonumber \\
&& + \frac{1}{4k} g_{\mu \nu} g^{\alpha \beta} \Bigg( \Big[ \left( \partial_\sigma {K_{\alpha \beta}}^\sigma - \partial_\alpha {K_{\sigma \beta}}^\sigma + {K_{\alpha \beta}}^\epsilon {K_{\delta \epsilon}}^\delta - {K_{\delta \beta}}^\epsilon {K_{\alpha \epsilon}}^\delta \right)
+ \left( \partial_\sigma {K_{\beta\alpha}}^\sigma - \partial_\beta {K_{\sigma \alpha}}^\sigma + {K_{\beta\alpha}}^\epsilon {K_{\delta \epsilon}}^\delta - {K_{\delta \alpha}}^\epsilon {K_{\beta \epsilon}}^\delta \right) \Big] \nonumber \\
&& + \left[ \left( \Gamma^{\epsilon}_{\,\,\,\alpha \beta} {K_{\delta \epsilon}}^\delta + {K_{\alpha \beta}}^\epsilon \Gamma^{\delta}_{\,\,\,\delta \epsilon} - \Gamma^{\epsilon}_{\,\,\,\delta \beta} {K_{\alpha \epsilon}}^\delta - {K_{\delta \beta}}^\epsilon \Gamma^{\delta}_{\,\,\,\alpha \epsilon} \right)
+ \left( \Gamma^{\epsilon}_{\,\,\,\beta\alpha} {K_{\delta \epsilon}}^\delta + {K_{\beta\alpha}}^\epsilon \Gamma^{\delta}_{\,\,\,\delta \epsilon} - \Gamma^{\epsilon}_{\,\,\,\delta \alpha} {K_{\beta \epsilon}}^\delta - {K_{\delta \alpha}}^\epsilon \Gamma^{\delta}_{\,\,\,\beta \epsilon} \right) \right]  \Bigg) \ .
\end{eqnarray}
It is easy to check that for vanishing spin-current Einstein equations are recovered as expected. 
This form is mostly convenient for its use in the previously found Raychaudhuri equations for the expansion given their form functionally identical to those in purely metric geometries.

\section{Gravitational collapse in Einstein--Cartan theory}\label{sec:ECcoll}
Having derived the full set of Raychaudhuri equations in the presence of torsion, and recast the ECT field equation in a GR form which identifies a generalized SET, we can now discuss the possible fate of a gravitational collapse in this gravitational setting. To keep things simple and manageable at the analytical level, we adopt an OS collapse framework with collapsing mass and initial radius equal to those of a typical neutron star, say $M=2 \times M_{\odot} $ and $R_i = 10^4$ m, where $M_{\odot}$ is a solar mass. As said, in this model the collapsing matter is considered a homogeneous dust ball with pressure, $P=0$.
\subsection{Oppenheimer--Snyder Collapse}
Let us start by presenting a brief overview of the OS collapse geometry, more details can be found in the standard references~\cite{Misner1974,Poisson2009}. 
The line element of the OS collapse geometry reads
\begin{eqnarray} \label{eq:83}
ds^2 = -c^2 d \tau^2 + a^2(\tau) \left(d \chi^2 + \sin^2 \chi d \Omega^2 \right)\ ,
\end{eqnarray}
where $a(\tau)$ is the scale factor, $d \Omega^2 = d\theta^2 + \sin^2 \theta d\phi^2 $ and the coordinate $0 \leq \chi \leq \chi_0$ (the surface of the star located at $\chi=\chi_0$) is comoving with collapsing dust.
%
From Einstein's field equation and SET conservation we can write
\begin{eqnarray}
\dot{a}^2 + c^2 &=& \frac{8 \pi G}{3} \rho a^2 \ , \label{eq:84} \\
\rho_0 a^3 &=& \textrm{constant} = \frac{3 c^2}{8 \pi G} a_m \ ,\label{eq:85}
\end{eqnarray}
where $\rho_0$ is the mass-density measured by comoving observer and $a_m$ is the maximum value of the scale factor.
From Eqs.~\eqref{eq:84} and~\eqref{eq:85} we can write parametrically
\begin{eqnarray}
a &=& \frac{1}{2} a_m \left( 1 + \cos \eta \right) \ , \label{eq:86}\\
\tau &=& \frac{1}{2c} a_m \left( \eta + \sin \eta \right) \ . \label{eq:87}
\end{eqnarray} 
where $\eta$ is the conformal time defined as $d\eta=c d\tau/a$.
We see from Eq.~\eqref{eq:86} that the collapse begins at $\eta=0$ when scale factor is maximum i.e. $a=a_m$ and it ends at $\eta=\pi$ when scale factor is zero i.e.~$a=0$.
Matching the solution with Schwarzschild exterior one gets total mass and time dependent radius, 
\begin{eqnarray} \label{eq:88}
M = \frac{c^2}{2 G} a_m \sin^3 \chi_0 \ , \quad \textrm{and} \quad R(\tau) = a(\tau) \sin \chi_0 \ ,
\end{eqnarray}
where initial radius can be written from above as $R_i =R(0)= a_m \sin \chi_0$. 
From above expression, the initial as well as maximum values of the scale factor and the maximum value of $\chi$ coordinate can be written as
\begin{eqnarray}
a_m &=& \sqrt{\frac{c^2 R_i^3}{2 G M}} \ , \label{eq:89} \\
\chi_0 &=& \arcsin \sqrt{\frac{2GM}{c^2 R_i} } \label{eq:90} \ .
\end{eqnarray}
The density varies with proper time as
\begin{eqnarray} \label{eq:91}
\frac{\rho_0 (\tau)}{\rho_0 (0)} = \left( \frac{a_m}{a(\tau)} \right)^3 \ ,
\end{eqnarray}
where $\rho_0 (0) = 3M/4 \pi R_i^3$ is the density measured at $\tau = 0$ by comoving observer. Using \eqref{eq:86} in \eqref{eq:91} we can write the density as a function of the conformal time as
\begin{eqnarray} \label{eq:92}
\rho_0 (\eta) = \frac{3M}{4 \pi R_i^3} \sec^6 \left(\frac{\eta}{2}\right) \ .
\end{eqnarray}
Using Eqs.~\eqref{eq:87} and \eqref{eq:89}, the line element in \eqref{eq:83} can be rewritten as
\begin{eqnarray} \label{eq:93}
ds^2 = \frac{c^2 R_i^3}{2GM} \cos^4 \left( \frac{\eta}{2} \right) \left[ -d \eta^2 + d \chi^2 + \sin^2 \chi d\Omega^2 \right] \ .
\end{eqnarray}
So it will take a finite amount of proper time, $\tau (\pi) = \frac{\pi M}{\sqrt{G}} \left( \frac{R_i}{2M} \right)^{3/2}$ for each collapsing dust particle to reach the singularity.

\subsection{SET for spinning dust}
Now let us see how a SET can be defined for collapsing dust. We consider a dust distribution like Weyssenhoff~\cite{Weyssenhoff1947} consisting of particles having intrinsic spin angular momentum~\cite{Trautman2006} characterized by a stress-energy tensor and a spin-angular momentum tensor given by
\begin{eqnarray} \label{eq:94}
\Sigma_{\mu \nu} = p_\mu u_\nu  \quad \mathrm{and} \quad \tau_{\mu \nu \eta} = s_{\mu \nu} u_\eta \ ,
\end{eqnarray}
where $u_\nu$ is four velocity, $p_{\mu}$ is density of four momentum given by
%
\begin{equation} \label{eq:95}
p_\mu = \rho_0 u_\mu - \frac{1}{c^2} \dot{s}_{\mu \nu} u^\nu \ ,
\end{equation}
$s_{\mu \nu}$ is the tensor of spin angular momentum density of collapsing dust.

We consider the fluid at rest which implies
\begin{eqnarray} \label{eq:96}
u_{\nu} = (c\sqrt{-g_{\eta \eta}},0,0,0) \ .
\end{eqnarray}

The spin-angular momentum density tensor is an antisymmetric tensor satisfying
\begin{eqnarray} \label{eq:97}
s_{\mu \nu}+ s_{\nu \mu} = 0 \ .
\end{eqnarray}
%

%
%
\subsection{Spin angular momentum density tensor}
In what follows we shall assume that all the components of the spin angular momentum density are zero with the exception of
\begin{eqnarray}
 s_{\phi \chi}=-s_{\chi \phi} &=& \frac{a^2(\eta ) \sigma (\eta ) \sin \theta \tan \theta \sin \chi \tan \chi }{\sqrt{\tan ^2 \theta \tan ^2 \chi +\sin ^2 \chi}}  \ , \label{98} \\
 s_{\phi \theta}=-s_{\theta \phi} &=& \frac{a^2(\eta ) \sigma (\eta ) \sin \theta \sin ^3 \chi}{\sqrt{\tan ^2 \theta \tan ^2 \chi +\sin ^2 \chi }} \label{99} \ .
\end{eqnarray}
It is important to note that the Frenkel condition, $s_{\alpha \beta} u^\beta = 0$ holds which basically closes the system of Matthison-Papapetou equations~\cite{Papapetrou1951,Mathisson1937}.

\subsection{Effective SET}
Following the above considerations we can now derive for the effective SET~\eqref{eq:82} for our neutron star
\begin{eqnarray}
\Sigma^{\mathrm{eff}}_{\eta \eta} &=& \frac{c^4 R_i^3 \rho_0 (\eta ) \cos^4 \frac{\eta }{2}}{2 G M} - \frac{16 \pi  G^2 M a^4 (\eta ) \sigma^2 (\eta ) }{c^4 R_i^3 \cos^4 \frac{\eta }{2}} \ ,   \label{eq:100} \\
\Sigma^{\mathrm{eff}}_{\chi \chi} &=& - \frac{16 \pi  G^2 M a^4 (\eta ) \sigma^2 (\eta )}{c^4 R_i^3 \cos^4 \frac{\eta }{2}} \ , \label{eq:101} \\
\Sigma^{\mathrm{eff}}_{\theta \theta} &=& - \frac{16 \pi  G^2 M a^4 (\eta ) \sigma^2 (\eta ) \sin^2 \chi  }{c^4 R_i^3 \cos^4 \frac{\eta }{2}} \ , \label{eq:102} \\
\Sigma^{\mathrm{eff}}_{\phi \phi} &=& - \frac{16 \pi  G^2 M a^4 (\eta ) \sigma^2 (\eta )  \sin^2 \theta \sin^2 \chi }{c^4 R_i^3 \cos^4 \frac{\eta }{2}} \ . \label{eq:103}
\end{eqnarray}
\subsection{Spin contribution to the effective SET}
The spin density in the above expression can be established as
\begin{eqnarray} \label{eq:104}
\sigma(\eta) &=& \gamma \times \frac{\hbar \rho_0(\eta)}{2 m_n} \ ,
\end{eqnarray}
where $\gamma$ is spin-alignment parameter determined by amount of alignment of the neutron's spin~(For example, among ten neutrons one is along negative z-axis and rest of the neutrons are along positive z-axis, in this case the value of gamma will be $0.8$. So gamma basically determines effective spin per neutron.), $m_n$ is the mass of the neutron and $\rho_0(\eta)$ is given by \eqref{eq:92}. Let us now check when the spin contribution to the SET is of the same order as the pure dust contribution. From~\eqref{eq:100}, we can separate the pure dust and torsion part as,
\begin{eqnarray} \label{eq:105}
\Sigma^{\mathrm{dust}}_{\eta \eta} = \frac{c^4 R_i^3 \rho_0 (\eta ) \cos^4 \frac{\eta }{2}}{2 G M} \ , \quad 
\Sigma^{\mathrm{spin}}_{\eta \eta} = - \frac{16 \pi  G^2 M a^4 (\eta ) \sigma^2 (\eta ) }{c^4 R_i^3 \cos^4 \frac{\eta }{2}} \ .
\end{eqnarray}
The absolute value of the ratio of the spin contribution to the dust one is then
\begin{eqnarray} \label{eq:106}
\displaystyle\left\lvert \frac{\Sigma^{\mathrm{spin}}_{\eta \eta}}{\Sigma^{\mathrm{dust}}_{\eta \eta}} \right\rvert &=& \frac{32 \pi  G^3 M^2 a^4 (\eta ) \sigma^2 (\eta ) }{c^8 R_i^6 \rho_0 (\eta ) \cos ^8 \frac{\eta }{2}} \ .
\end{eqnarray}
%
%
%
Using~\eqref{eq:86}, \eqref{eq:88}, \eqref{eq:104} and plugging the value of all quantities~(see Appendix~\ref{app:num}) we get
\begin{eqnarray} \label{eq:107}
\displaystyle\left\lvert  \frac{\Sigma^{\mathrm{spin}}_{\eta \eta}}{\Sigma^{\mathrm{dust}}_{\eta \eta}} \right\rvert = \frac{6 \gamma^2 G^3 M^3 a_m^7 \hbar^2 \sin ^3 \chi_0}{c^8 R_i^9 m_n^2 R^3} \approx \frac{2 \times 10^{-28} \gamma^2}{ R^3}  \ .
\end{eqnarray}
When the spin contribution is comparable to the pure dust contribution then the ratio in~\eqref{eq:107} approaches to unity. 
We can then compute the radius~$R(\gamma)$, at which this ratio becomes unity, for various values of spin-alignment~$\gamma$.
As a matter of fact such radius turns out to be weakly sensitive to the strength of spin-alignment parameter so that for example $R(\gamma)$ ranges from $R(0.2) = 1.99 \times 10^{-10} $ m to $R(1) = 5.81 \times 10^{-10}$ m. It is worth noticing that this radius is way smaller that the one at which a trapping horizon form $R=2GM\sim 6$ km 
meaning that torsion cannot prevent the formation of the latter. Nonetheless, the equality between spin and dust contribution equality is reached also well before the onset of Planck densities $\rho_{\rm Pl} = M_{\rm Pl}/V_{\rm Pl}\approx 10^{95}$ kg/$m^3$  which for our neutron star are reached for $R\approx 10^{-23}$ m.

%
\subsection{Null-Energy Condition}
A general null four vector can be written as
\begin{equation} \label{eq:108}
k^{\mu} = \left(1, \sqrt{\frac{-g_{\eta \eta}}{g_{\chi \chi}}},0,0 \right)= (1,1,0,0) \ .
\end{equation}

Using Eqs.~\eqref{eq:100}-\eqref{eq:103} and~\eqref{eq:108}, the null energy condition reads
\begin{equation} \label{eq:109}
\Sigma^{\mathrm{eff}}_{\mu\nu} k^\mu k^\nu = \frac{c^4 R_i^3 \rho_0 (\eta ) \cos ^4 \frac{\eta }{2}}{2 G M}-\frac{32 \pi  G^2 M a^4 (\eta ) \sigma^2 (\eta ) }{c^4 R_i^3 \cos ^4 \frac{\eta }{2}} \geq 0 \ .
\end{equation}
Our main goal is to check if there will be violation of NEC at any stage of collapse. Using~\eqref{eq:86}, \eqref{eq:88} ,\eqref{eq:104} and putting all numerical values~(as of Appendix~\ref{app:num}) we get
\begin{eqnarray} \label{eq:110}
\Sigma^{\mathrm{eff}}_{\mu\nu} k^\mu k^\nu = \frac{3 c^4 a_m \sin \chi_0}{8 \pi G R} - \frac{9 \gamma^2 G^2 M^3 a_m^8 \hbar^2 \sin ^4 \chi_0 }{2 \pi c^4 R_i^9 m_n^2 R^4} \approx \frac{1.45 \times 10^{47}}{R}-\frac{5.68 \times 10^{19} \gamma ^2}{R^4} \ .
\end{eqnarray}
If $\Sigma^{\mathrm{eff}}_{\mu\nu} k^\mu k^\nu$ is negative then NEC will be violated. 
This again can happen for $\eta \rightarrow \pi$, indeed one can easily find the radius~$R(\gamma)$ of the configuration for various values of spin-alignment parameter~$\gamma$ below which NEC will be violated.
The result is again relatively insensitive to the strength of the spin alignment so that $R(\gamma)$ ranges from $R(0.2) = 2.5 \times 10^{-10}$ m to $R(1)=7.32 \times 10^{-10}$ m.
It can be shown that for similar $R(\gamma)=O(10^{-10})$ m all the other energy conditions are also violated (see Appendix~\ref{app:EC}).

\subsection{Avoidance of singularity}
We just saw that the equivalence of the spin and dust components of the SET happens basically at the same radius at which the NEC is violated. This is not so surprising as the spin component is always NEC violating and more so as the radius shrinks. Let us stress again that this happens for radii of order $R=O(10^{-10})$ m which are small but still larger than the scale at which the stars density becomes Planckian $R=O(10^{-23})$ m (see Appendix~\ref{app:PlanckD}) or even more of the order of the Planck length $R=O(10^{-35})$ m. This implies that one of the assumptions, on which Penrose's singularity theorem rests, will break down in the late stages of our collapse model before reaching a quantum gravity regime due to the presence of torsion sourced by the spin current in ECT. 

In order to be more concrete, let us consider how the violation of the NEC entails the removal of the starting point of Penrose's theorem, i.e.~the presence of a focusing points for the null geodesic of the collapse geometry. The Raychaudhuri equation for the expansion of null geodesics congruences is given in~\eqref{eq:57}. Specializing to our spherically symmetric OS collapse it becomes
%
\begin{eqnarray} \label{eq:111}
\frac{D \theta}{d \lambda}  &=& 
- \frac{1}{2} \theta^2 -\widetilde{R}_{\eta \rho} k^\eta k^\rho\ .
\end{eqnarray}
Using~\eqref{eq:81} and~\eqref{eq:110}, we can write
\begin{eqnarray} \label{eq:112}
\widetilde{R}_{\eta \rho} k^\eta k^\rho = k \Sigma^{\mathrm{eff}}_{\mu\nu} k^\mu k^\nu = \frac{3 c^4 a_m \sin \chi_0}{c^4 R} - \frac{36 \gamma^2 G^3 M^3 a_m^8 \hbar^2 \sin ^4 \chi_0 }{2 \pi c^8 R_i^9 m_n^2 R^4}  \ .
\end{eqnarray}
We can write the expression of expansion parameter~($\theta$) as follows
\begin{eqnarray} \label{eq:113}
\theta = \left(2 \cot \chi -2 \sqrt{\frac{R_i}{R}-1} \right) \ .
\end{eqnarray}
One can clearly see from expression~\eqref{eq:113} that expansion diverges at $\chi=0$. This is an artifact of OS model.
Using~\eqref{eq:112}, \eqref{eq:113} and putting all numerical values~(as of Appendix~\ref{app:num}) we get
\begin{eqnarray}
\frac{D \theta}{d \lambda}  &=& - \frac{1}{2} \left(2 \cot \chi -2 \sqrt{\frac{R_i}{R}-1}\right)^2 - \frac{3 c^4 a_m \sin \chi_0}{c^4 R} + \frac{36 \gamma^2 G^3 M^3 a_m^8 \hbar^2 \sin ^4 \chi_0 }{2 \pi c^8 R_i^9 m_n^2 R^4} \nonumber \\
&\approx& \frac{1}{2} \left(2 \cot \chi -2 \sqrt{\frac{R_i}{R}-1}\right)^2 - \frac{3 \times 10^{4}}{R} + \frac{1.18 \times 10^{-23} \gamma ^2}{R^4}
\end{eqnarray}
To provide an example, let us investigate the bounce from above equation at star's surface i.e. $\chi=\chi_0$ and when spins are fully aligned i.e. $\gamma=1$, the radius at which we see a bounce is $R= 5.52 \times 10^{-10}$ m.
%
%
%
%

%

%

\section{Discussion}\label{sec:disc}
We have derived here the modified Raychaudhuri equations in a spacetime with torsion for expansion, shear and vorticity considering both timelike and null congruences. 
Remarkably, the form of Raychaudhuri equation for the expansion is formally the same as for purely metric geometries for both timelike and null congruences~\footnote{While our results are in accordance with those of ref.~\cite{Luz2017}, to our knowledge the timelike case was not analyzed before.}. However, it is important to stress that in both cases (see Eqs.~\eqref{eq:33} and \eqref{eq:57}) the Ricci tensor is not same as it would be without torsion, and indeed it is rather given by Eq.~\eqref{eq:79}. So while the Raychaudhuri equation for the expansion does not formally change in a spacetime with torsion, still torsion affects it.

We then considered an OS collapse model which assumes the ideal case of the collapse of a star made of pressureless ($P=0$) dust. 
In addition, we considered the collapsing dust being composed of particles with intrinsic spin (so to have a source for torsion as e.g.~in \cite{Trautman1973}) and calculated the effective energy-momentum tensor given in Eq.~\eqref{eq:82}. While spin alignment could be expected in the presence of large magnetic fields such as those characterizing neutron stars, we took into account some degree of disruption of the spin alignment by introducing a suitable parameter (to which nonetheless our results resulted to be weakly sensitive).

Although one could argue that inside typical neutron stars the density is very high (even more so in the late stages of the collapse after the horizon crossing), so that considering the OS solution may not be appropriate, more realistic forms of matter would add pressures which however, due to the much faster growth of the spin component of the stress energy tensor for small radii (see e.g.~Eq.~\eqref{eq:107} or Eq.~\eqref{eq:110}) would not radically change our conclusions. Hence, this toy model has the merit to be easily manageable and to be able to provide a physical intuition about the role that spin currents and the associated torsion could have in the late phases of a stellar collapse after the formation of a horizon.

At least in the present scenario considering a two solar masses collapsing star, it is found that the torsion contribution in the stress-energy tensor is comparable to the dust one when the radius is of order $\sim 10^{-10}$ m. At roughly the same radius all the energy conditions are also violated. We saw that this radius is much bigger than the radius at which the density becomes Planckian,  $\sim 10^{-23}$ m, and hence one would need a full quantum gravity treatment. This indicates that if the constituent particles of the collapsing dust star have an intrinsic spin then at some point of the collapse the discrepancies between Einstein-Cartan theory and Einstein gravity would become evident and due to this the formation of a singularity could in principle be avoided, possibly leading to some form of regularized black hole interior~\cite{Carballo-Rubio2019a,Carballo-Rubio2019b}. This conclusion seems also to lend support to the outcome of the numerical investigation carried out in~\cite{Hashemi2015}. There an OS collapse in ECT was also considered and it was found, via a numerical analysis, that the singularity formation is resolved by a bouncing geometry (another typical scenario in quantum gravity settings~\cite{Barcelo2014a,Haggard2014,Malafarina:2017csn,Bianchi:2018}). We hope to further explore this possibility in future work.

\section*{Acknowledgements}
The authors would like to thank Enrico Barausse, Vincenzo Vitagliano and Edgardo Franzin for useful discussions. SH would like to acknowledge the institutional support of Institute of Physics, Silesian University in Opava and the internal student grant SGS/12/2019 of Silesian University in Opava. SH would also like to thank SISSA for the hospitality during the early phase of this work. SL acknowledges funding from the Italian Ministry of Education and Scientific Research (MIUR) under the grant PRIN MIUR 2017-
MB8AEZ.
\appendix

\section{Numerical values of parameters/constants} \label{app:num}
\begin{eqnarray} 
\textrm{Gravitational constant, } G &=& 6.67408 \times 10^{-11} \mathrm{m^3 \cdot kg^{-1} \cdot s^{-2}} \nonumber \\
\textrm{Velocity of light, } c &=& 3 \times 10^8  \mathrm{m \cdot s^{-1}} \nonumber \\
\textrm{Reduced Planck constant, } \hbar &=& \frac{6.626 \times 10^{-34}}{2 \pi} \mathrm{kg \cdot m^2 \cdot s^{-1}  (J \cdot s)}\nonumber \\
\textrm{Mass of the neutron, } m_n &=& 1.6749 \times 10^{-27} \mathrm{kg} \nonumber \\
\textrm{Solar mass, } M_{\odot} &=& 2 \times 10^{30} \mathrm{kg} \nonumber \\
\textrm{Mass of the neutron star, } M &=& 2 \times M_{\odot} \nonumber \\
\textrm{Initial radius of the neutron star, } R_0 &=& 10^4 \mathrm{m}  \nonumber \\
\end{eqnarray}

\section{Planck density}
\label{app:PlanckD}
Let us consider that at $\eta=\eta_*$ the density of the distribution is equal to  Planck density, $\rho_{_{\textrm{Planck}}}=10^{96}$ $\mathrm{kg \cdot m^{-3}}$.
\begin{eqnarray} \label{C1}
\rho_0 (\eta_*) &=& \rho_{_{Planck}} \nonumber \\
\frac{3 M}{4 \pi R_0^3} \cos^{-6} \left(\frac{\eta_*}{2}\right) &=& 10^{96} \nonumber \\
\cos^{6} \left(\frac{\eta_*}{2}\right)  &=& \frac{3 M}{4 \pi R_0^3}\times 10^{-96} \nonumber \\
\cos \left(\frac{\eta_*}{2}\right)  &=& \left( \frac{3 M}{4 \pi R_0^3}\times 10^{-96} \right)^{1/6 } \nonumber \\
\eta_* &=& 2 \arccos \left( \frac{3 M}{4 \pi R_0^3}\times 10^{-96} \right)^{1/6 } .
\end{eqnarray}
The density of the distribution in our consideration reaches to Planck density at $\eta_*=0.999999999999937 \hspace{0.02cm} \pi$ and then scale factor is $\sim 10^{-22}$ \quad m. We can also find the actual radius of the dust cloud is $\sim 10^{-23}$ m.

\section{Other Energy Conditions}
\label{app:EC}

\subsection{WEC}
WEC states that 
\begin{eqnarray}
\Sigma^{\mathrm{eff}}_{\mu\nu} u^\mu u^\nu \geq 0  \quad \forall u^\mu \quad \mathrm{timelike} \ ,
\end{eqnarray}
where in our consideration $u^\nu$ is given by Eq.~\eqref{eq:96}.
Using Eqs.~\eqref{eq:96} and~\eqref{eq:100}-\eqref{eq:103}, the weak energy condition takes the form
\begin{eqnarray}
\Sigma^{\mathrm{eff}}_{\mu\nu} u^\mu u^\nu = c^4 \rho_0 (\eta )-\frac{32 \pi  G^3 M^2 a^4(\eta ) \sigma^2 (\eta ) }{c^4 R_i^6 \cos ^8 \frac{\eta }{2} } \geq 0 \ .
\end{eqnarray}
Plugging all numerical values~(as of~\eqref{app:num}) we get
\begin{eqnarray}
\Sigma^{\mathrm{eff}}_{\mu\nu} u^\mu u^\nu = \frac{3 c^4 M a_m^3 \sin^3 \chi_0}{4 \pi R_i^3 R^3} - \frac{9 \gamma^2 G^3 M^4 a_m^{10} \hbar^2 \sin ^6 \chi_0 }{2 \pi c^4 R_i^{12} m_n^2 R^6} \approx \frac{7.73 \times 10^{63}}{R^3} -\frac{1.51 \times 10^{36} \gamma ^2}{R^6} \ .
\end{eqnarray}
If $\Sigma^{\mathrm{eff}}_{\mu\nu} u^\mu u^\nu$ is negative then WEC will be violated.
By similar treatment like we did using \eqref{eq:110}, one gets the radius~($R(\gamma)$) of the configuration below  which WEC violation occurs ranging from $R(0.2) = 1.99 \times 10^{-10} $ m to $R(1) = 5.81 \times 10^{-10}$ m .
%
\subsection{DEC}
Using Eq.~\eqref{eq:96}, the flux measured by an observer along the congruence is given by
\begin{eqnarray}
F^\mu = \Sigma^{\mathrm{eff}}_{\mu\nu} u^\nu = \frac{ \cos ^2 \frac{\eta }{2} \left(32 \pi  G^3 M^2 a^4(\eta ) \sigma^2 (\eta ) \sec ^8 \frac{\eta }{2} -c^8 R_i^6 \rho_0 (\eta )\right)}{\sqrt{2GM} c^4 R_i^{9/2}} \ .
\end{eqnarray}
DEC states that WEC holds plus the above flux vector is not spacelike for any timelike observer $u^\nu $ i.e.
\begin{eqnarray}
F^\mu F_\mu \leq 0 \ .
\end{eqnarray}
We find the norm of the flux vector
\begin{eqnarray}
F^\mu F_\mu = -\frac{\left(c^8 R_i^6 \rho_0 (\eta )-32 \pi  G^3 M^2 a^4 (\eta ) \sigma^2 (\eta ) \sec ^8 \frac{\eta }{2}\right) ^2}{c^{10} R_i^{12}} \ .
\end{eqnarray}
In our case, we clearly see that flux is always timelike which means that there is no superluminal flux. But as WEC is violated at late time of collapse ($\eta \rightarrow \pi$) so does DEC.
\subsection{SEC}
SEC requires that
\begin{eqnarray}
\left(\Sigma^{\mathrm{eff}}_{\mu\nu} - \frac{1}{2} \Sigma^{\mathrm{eff}} g_{\mu \nu} \right) u^\mu u^\nu \geq 0 \quad \forall u^\mu \quad \mathrm{timelike} \ ,
\end{eqnarray}
where $\Sigma^{\mathrm{eff}} = g^{\mu \nu} \Sigma^{\mathrm{eff}}_{\mu\nu}$.
Using Eqs.~\eqref{eq:96} and~\eqref{eq:100}-\eqref{eq:103}, the strong energy condition reads
\begin{eqnarray}
\left(\Sigma^{\mathrm{eff}}_{\mu\nu} - \frac{1}{2} \Sigma^{\mathrm{eff}} g_{\mu \nu} \right) u^\mu u^\nu = \frac{c^4 \rho_0 (\eta )}{2}  -\frac{64 \pi  G^3 M^2 a^4 (\eta ) \sigma^2 (\eta ) }{c^4 R_i^6 \cos ^8 \frac{\eta }{2}} \geq 0 \ .
\end{eqnarray}
Putting all values (as of \eqref{app:num}) we get
\begin{eqnarray}
\left(\Sigma^{\mathrm{eff}}_{\mu\nu} - \frac{1}{2} \Sigma^{\mathrm{eff}} g_{\mu \nu} \right) u^\mu u^\nu = \frac{3 c^4 M a_m^3 \sin^3 \chi_0}{8 \pi R_i^3 R^3} - \frac{9 \gamma^2 G^3 M^4 a_m^{10} \hbar^2 \sin ^6 \chi_0 }{\pi c^4 R_i^{12} m_n^2 R^6} \approx \frac{3.87 \times 10^{63} }{R^3} - \frac{3.03 \times 10^{36} \gamma ^2}{R^6} \ .
\end{eqnarray}
We also see at late time of collapse ($\eta \rightarrow \pi$) SEC is violated. Again, we compute the radius~$R(\gamma)$ of the configuration below  which above expression becomes negative i.e. violation of SEC occurs, indeed one can easily find $R(\gamma)$ ranging from $R(0.2) = 3.15 \times 10^{-10} $ m to $R(1) = 9.22 \times 10^{-10} $ m .

\bibliography{raychmain}
\bibliographystyle{ieeetr}

\end{document}